# Why are we not able to see beyond three dimensions?

Rogério Martins[1]
Centro de Matemática e Aplicações (CMA), FCT, UNL
Departamento de Matemática, FCT, UNL
Faculdade de Ciências e Tecnologia da Universidade Nova de Lisboa

Abstract: This is perhaps a philosophical question rather than a mathematical one, we do not expect to give a full answer, even though we hope to clarify some ideas. In addition, we would like to provide a new perspective on the subject. We will find curious analogies with the way we perceive color and make some imaginary experiments showing that, even living imprisoned in three dimensions… it could be different.

———

Did you ever see a hypersphere? I guess not. So… why not? This is the question we all, sooner or later have posed to ourselves. Why do we have this blockade when passing from three to higher dimensions? Could this have been different? Is there hope that in the future we can overcome this condition?

Let us start by giving an attempt to clarify what we usually mean when we say that we are "seeing" or "visualizing" a geometrical object, for example a sphere.

On one hand, there is the pure mathematical object, that we all know as the sphere, this sphere has a mathematical characterization and lives in an abstract space, in the Platonic sense if you will. On the other hand, there is our physical experience of a real sphere, something that we can perceive with our senses and spacial intuition. What do we mean by "seeing" a mathematical sphere? It is, in my opinion, this possibility to imagine the mathematical sphere in our three dimensional physical world, something that could have been real, like a soccer ball. Even as a product of our imagination, we can imagine some physical interaction with it, holding it, rotating it or changing its position. Somehow, we can use our three dimensional physical intuition to understand the sphere's properties and the way it interacts with other geometrical objects. We use our three dimensional physical space to understand the abstract Euclidean three dimensional vector space.

[1] This work was partially supported by the Fundação para a Ciência e a Tecnologia (Portuguese Foundation for Science and Technology) through the project UID/MAT/00297/2013 (Centro de Matemática e Aplicações)

It's funny to think this way, certainly the idea of a vector space, and the sphere, was created upon the physical sensible experience of space. Anyway, the abstract idea of vector space and its geometric objects gained a life of their own and its properties were generalized to higher dimensional spaces. Now we have a fourth dimensional object, that we call an hypersphere, with similar properties to the three dimensional counterpart, and we do not have the correspondent physical object to "see" it.

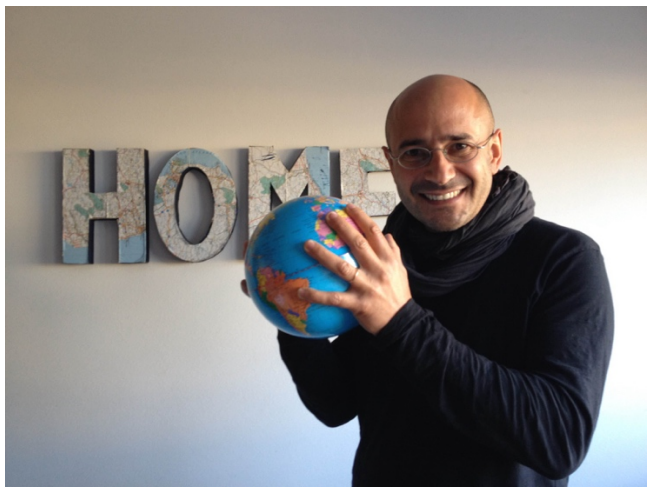

*Figure 1: We can use our three dimensional physical intuition to understand the mathematical sphere's properties.*

Recent evidence from neurobiology shows that our brain seems to have been originally created to manage our homeostatic mechanisms and physical interaction with our surroundings [5] [6]. Memory, conscience, planning and deciding seems to be superimposed on the top of a cerebral structure originally build to deal with sensation and movement. Almost every part of our brain, even being responsible for some other cognitive function, has some sensory and motor signals. Apparently when we imagine a physical sensation, for example holding a sphere, we are partially activating the very same tactile responses that would be activated if the sphere were really in our hands [7][8]. These ideas are also somehow connected with a philosophical theory called *embodied cognition*.

Therefore, our physical sensible intuition must be incomparably stronger than our mathematical intuition in general, this is probably what gives us this sensation of "seeing" a sphere in opposition to comprehend a sphere, when we "see" we are using this strong spacial intuition. This is probably why we feel this powerlessness when moving from three to four dimensions.

This is usually the end of the story, we cannot see an hypersphere simply because we live in a three dimensional physical space. We will try to convince you that it could be different.

This is of course a question of mathematical relevance; this limits our daily mathematical activity. So, why not give it a try? At least we hope to clarify some ideas. With this in mind we will try to find out some analogies with the way we perceive color and make some imaginary experiments that we hope, will convince you that, even imprisoned in a three dimensional physical space, we could have been different.

We propose that this limitation is simply a characteristic of our species, it is given by our biology, that in turn was shaped in order to succeed in our environment, from an evolutionary point of view. A different sensory system or body characteristics could had equipped us with other perception capabilities of the mathematical geometric objects.

**The higher dimensions**

We all remember the adventures of the square that inhabited the world of Edwin Abbott Abbott's novel, *Flatland: A Romance of many dimensions* [3][4]. This square, confined to its two dimensional world, cannot imagine what the third dimension is, until it establishes some connection with a sphere that showed him the third dimension. This sphere not only offers the square an excursion through the three dimensional space but also calls the attention for some relations between the two and three dimensions, that give us, inhabitants of a three dimensional space, some intuition on the fourth dimension. This leaves us humans, wondering if, like the square, we really live in a subspace of a fourth dimensional world, and how we perceive a fourth dimensional object that crosses our world.

There are of course several tricks to apprehend those higher dimensions. Some advantage can be taken from the temporal dimension, the use of projections can also be helpful, analogies with smaller dimensions, and so on. It is worth to refer the work of Charles Howard Hinton in this direction. However, this does not give a satisfactory solution, we still cannot see the hypersphere the same way we see the sphere. Of course that a fairly good comprehension of these higher dimensional objects can be attained, understanding their properties, and manipulating them. But, this does not mean that we have a clear visualization of those structures, similar to visualization we have of the correspondent three dimensional ones: comprehend is different from "seeing" it. Very often we have no much more than intuition based on the lower dimensional version that we can "see."

All these dimensional questions have become recurrent in several areas apart from mathematics. The fiction industry, cinema, theatre, and literature, recurrently cast the issue. Pseudoscience also uses the ideas of hidden dimensions to base a bunch of ideas. On the other hand, the question of the number of dimensions is raised in quantum physics's last theory to explain the sub-atomic world, and show that after all we live in a ten, or so, dimensional world. However, even if a ten dimensional string theory is very adequate to give a mathematical model to the world we live in, the only things these extra dimensions have in common with the three spacial dimensions we perceive its the name and some mathematical similarities, indeed is like comparing chalk to cheese.

In this paper we do not want to speculate about the existence of extra dimensions that we cannot "see". The problem is a bit different, we are concerned with the problem that every student faces when learning higher dimensional mathematics, they feel that there is a limitation on what they can visualize. Surprisingly, almost nothing have been written about it.

**The way we perceive color**

When we were kids someone told us that every color could be obtained from the mixing of three colors, the primaries ones. Let's assume we are dealing with an additive color system, for example beams of light, the primaries commonly used are red, green and blue, all the other ones can be produced as a combination of different intensities of these three colors.

Indeed, experimental evidence shows that, while three colors is enough, no set of two colors can be mixed to produce the complete palette we perceive. Since every color we can distinguish is made of a certain combination of intensities of these three colors, we can expect to arrange the full palette of colors in a three dimensional vector space. The set of primary colors being a basis of this space, multiply by a scalar corresponding to the application of a certain intensity to a certain color and summation of two colors corresponds to the mixture of these colors. As long as we keep ourselves far from extreme values of luminosities, extreme dim or bright lights, it turns out that our perception of color is consistent with this three dimensional vector space model. Actually the positive octant of a three dimensional vector space, since there is no real interpretation for multiplication of a color by a negative number. This is essentially the RGB color model.

However, this is just the way we perceive color, the reality is a bit different. In reality there is no pure beam of light, with just one electromagnetic frequency. Each beam of light is compounded of a mixture of wavelengths and can be described by a wavelength density distribution function defined over the visible spectrum. The set of real beams of light can be put in correspondence with the positive real continuous functions defined over an interval of the real line. Mathematically, the set of real beams of light is an infinite dimensional vector space.

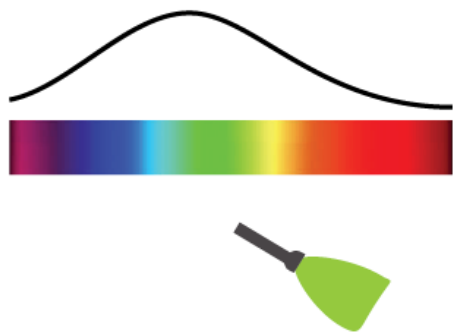

*Figure 2. Each beam of light is compounded of a mixture of wavelengths and can be described by a wavelength density distribution function defined on the visible spectrum.*

When a beam of light reaches our retina, it stimulates a certain group of cells, the so-called cone cells. An healthy person has three types of cone cells, each type is more sensitive to certain frequencies. These peaks of sensitivity do not correspond to the frequencies of the primary colors we use, however the fact that we use only three primary colors is a consequence of the existence of only three types of cone cells. Each beam of light stimulates each type to a certain degree, so our eyes eventually retain three values from a real beam of light and it is from those three values that our brain associates an idea of a certain color.

The fact that we, humans, have three types of cone cells and live in a three dimensional vector space seems to be a coincidence, actually in the animal world there is some diversity concerning the number of types of cone cells: it seems to be an evolutive feature of the species. The kangaroos and honeybees also have three types of cone cells, we are the so-called trichromats. Most of non-primate mammals are dichromats, only have two types of cone cells, this is also the case of some color blind humans. A dichromate only needs two primary colors to build its two dimensional chromatic space. Monochromacy is very rare among humans but this is the case of marine mammals and some sea lions, they only have one sensor, their chromatic space is unidimensional. On the other extreme there are the tetrachromats, with four types of cone cells, this is the case of some reptiles, birds, insects… and some women [1]. The fact that among humans only women have this characteristic is related to the fact that the genetic information concerning this cone cells is contained in the X chromosome. There are some species of monkeys where the females have trichromatic vision while the males are dichromat. There are also pentachromats, some butterflies and pigeons, and other species with more than five sensors, however they become more and more rare.

Mathematically, the set of colors we can perceive can be seen as the image, over a three dimensional space, of a projection applied on this infinite dimensional space of the real colors. This is a real occurrence of the allegory of Plato's cave. As a consequence, in reality there are many more "colors" than we can distinguish, there are different wavelength distributions that we humans perceive as the same color, this is the so-called metamerism. In some sense we are all color blind, indeed there are different real beams of light that we cannot distinguish.

Even though there is some hope, in 2009 a group of scientists succeed in injecting a virus in the eyes of a dichromate monkey and transformed some of those cone cells [2]. Of course this is good enough, however the results overcame the expectations of the group, since the brain of this adult monkey somehow managed to deal with the additional information sent by the modified cone cells, something, they believe, could only be possible in early stages of the development of the brain. This was confirmed since the monkey was trained to identify color spots on a screen, this monkey really started to distinguish colors that he did not differentiate before the operation.

This experience raises an interesting thought: what did this monkey feel when he started to see the new colors? In the Fig. 3., on the right, we see an image the way it is seen by a red-green color blind: the case exhibited by this monkey. On the left, we see the image as seen by a trichromat. Suddenly this monkey started to distinguish colors he did not experience before. It would be impossible to explain to this monkey, before the operation, how the color of the t'shirt was different from the green color of the board. This is probably one of the most similar sensations we can think of when seeing a fourth dimension.

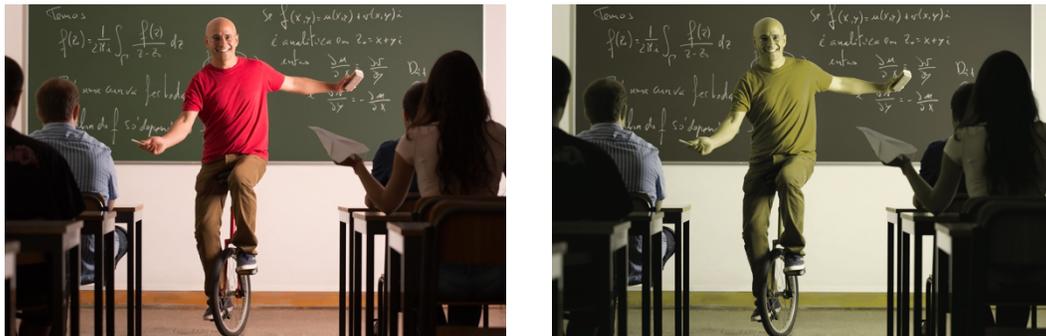

*Figure 3. A simulation of an image seen with normal color vision on the left and red-green color blindness on the right. Photo by Rodrigo de Matos.*

This experiment gives some hope to color blind people, this group of scientists believe that this open new possibilities in the treatment of color blindness in humans. Who knows if the same technology could also be used to create a fourth cone cell in a trichromat. Perhaps in the future we can ask for an extra sensor with the same easiness we ask to get a tattoo.

**A bunch of ideas raised by the color example**

The example above is interesting in its own, but it is also useful to clarify the problem we are trying to address. It raises some questions and allows some curious analogies. So, let us try to answer some:

Are there other options for a familiar vector space where we can "see"?

In the beginning of this paper we saw how do we use our three dimensional physical space to see or visualize a geometric object, for example a sphere, that lives in a three dimensional abstract vector space. We take advantage of our strong spatial intuition to understand the properties of this object, when we imagine this object as something real that we can feel. What if there is another intuitive vector space that we can use to visualize geometric objects?

Imagine someone who has such a complete mastery over color that he can imagine geometric objects in his chromatic space. For example, the Fig. 4. represents the set of colors that form a cube in our three dimensional chromatic space. We are also representing this set of colors in a picture that represent each color at a point in a three dimensional physical space. However, the idea is to conceive someone who has a so deep control over color that he does not feel the need to arrange the colors as points in a three dimensional physical space. Even though, this person has a clear idea of the properties of the vector space satisfied by the colors (points), which colors are in a neighborhood of a certain color, the distance between colors in its chromatic vector space, the result of the 'sum' of two colors, multiplication by a scalar and so on. This person could imagine directly the set of colors that form this cube and how these colors set out in his chromatic space.

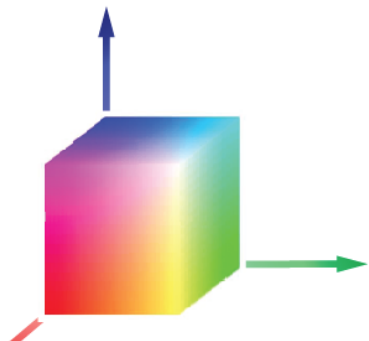

*Figure 4. The set of colors that form a tridimensional cube in the RGB color system.*

In principle this person can visualize any geometrical object in this space. However, since the chromatic space of a normal person is three dimensional, the same blockade happens in passing from the third to the fourth dimension. But maybe some tetrachromat woman wants to give it a try, or maybe in the future one of us can do the same with artificial extra cone cells.

Could it be different even living in a three dimensional space?

The color example shows us that our brain seems to be formatted depending on the way we interact with reality, each brain is adapted to a certain number of types of cone cells. In the same way, as we saw in the introduction, our brain was originally built essentially to deal with our physical interaction with physical surroundings, and so it was formatted accordingly.

We intuitively split and group our physical state in a spatial position, plus an direction of the body in this physical space, plus a position in time, and many other measures, like thermal sensation or noise level (Fig. 5.). For us, this is the natural split, each of these characteristics have their own physical and mathematical properties. The three spatial coordinates are similar in nature and a vector space. Then, direction of the body is no longer a vector space, mathematically it is a group, the usual SO(3), the group of all rotations about the origin. Time has different physical properties, and so one and so forth.

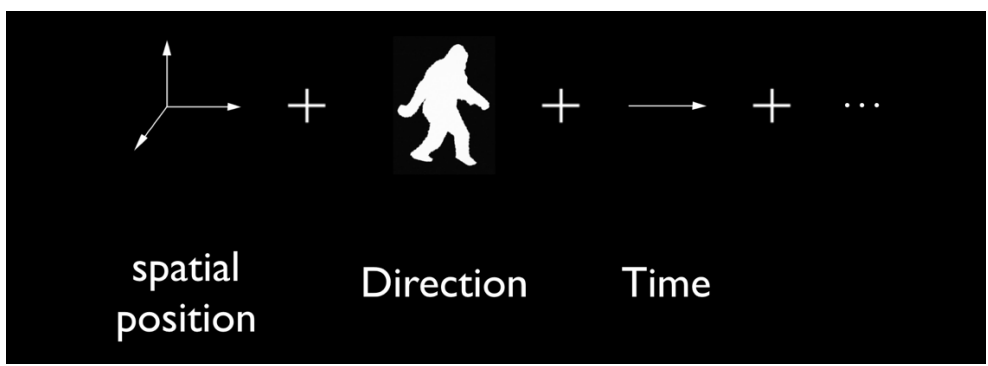

*Figure 5. We split and group our physical state in a tridimensional spacial position, plus an orientation in this physical space, plus a position in time, and so on.*

However, it could be different. Imagine that we were an animal that only moves in a relatively flat surface, say a kind of lizard in the desert, all our life happens in this plane, predators, food, mating, etcetera. It is not difficult to imagine that maybe in this case our brain cannot find a relation between the two planar coordinates, needed to position ourselves in this plane, and altitude.

Probably the orientation or time could be much more important for us than the altitude, that might not be even recognized (Fig. 6.). In this case, we would have a problem trying to see a sphere, probably this being would not see beyond two dimensions.

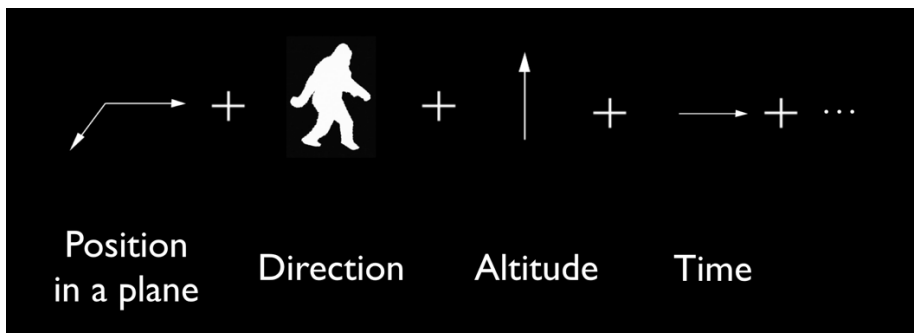

Figure 6. However, it could be different.

In a more extreme situation we can imagine that some brains can only be aware of one spatial dimension. Imagine for example lice, maybe the only important spatial data to take into account is the distance needed to reach the next head, an unidimensional quantity. If this brain attains a reasonable understanding of mathematics, then it cannot even "see" the circle.

What about seeing more than three dimensions?

Consider the following imaginary experiment: what if our body were made of two parts? Two parts but just one brain (Fig. 7). We are accustomed to the idea that to each body corresponds one brain, that in some cases can even have a conscience of itself. However, this is nothing but just one more characteristic of the species, like having two arms. I am not concerned with whether this make sense from the biological and evolutionary point of view. Surely there are some explanations missing. Would this brain be attached to just one part or would it be split? How is it linked to the parts? Nonetheless, do not worry about technical details and for the present moment just imagine a being made of one brain and a body separated into two disconnected parts.

Since each part of this body lives in our three dimensional physical world, it needs three spatial coordinates to specify its position. Eventually this brain could split the position of each part, say three dimensional position for part A plus three dimensional position for part B, however, the fact is that from a mathematical point of view, a six dimensional Euclidean vector space would be the right space to define the position of this being. Since this brain would have evolve together with this body, it is possible that it would deal directly with the six coordinates as a whole. In this case this being would had an intuitive six dimensional vector space to "see" the geometrical objects. Probably, if this being were the author of this paper the title would be: why are we not able to see beyond six dimensions?

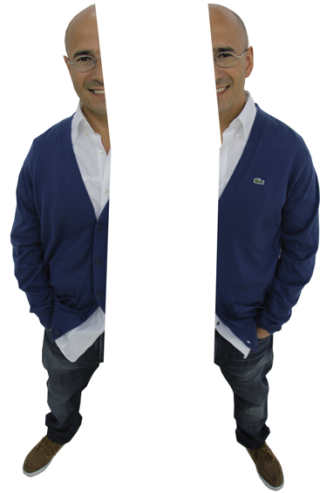

Figure 7. What if we were made of two parts?

Does the number of dimensions we perceive coincide with the number of real dimensions?

The last example showed us that we perceive a three dimensional chromatic space, even if the "real" chromatic space is better described as an infinite dimensional vector space. The first thought that comes to mind is that some similar phenomena could happen with our physical dimensions. Maybe we somehow only perceive three dimensions of a reality that is high dimensional, perhaps even infinite dimensional.

Well, in some sense, we really perceive a three dimensional physical space while there are higher dimensional mathematical models that better describe our reality, this is for example the case of the general theory of relativity. The fact that the "real" set of colors is modeled by an infinite dimensional space is, like the theory of relativity, just an abstract model that goes further in the description of our reality. The chromatic space generated by the three primaries colors or the three physical dimensions we perceive are a first approach based directly on our senses.

However, the higher dimensional mathematical models that describe the reality seem to show different qualitative properties among those dimensions. For example, the quantum theories, involve models with more dimensions, at any rate all this happens in scales where our senses are helpless. It would be difficult to compare the physical dimensions we perceive with the remaining ones. On the other hand, in the infinite dimensional model of color all the dimensions have a similar rule.

The idea of suddenly see a fourth spacial dimension certainly give a great storyline for a film or a book, however we do not know how it could make sense.

**Conclusion**

Finally, it seems that "seeing" up to three dimensions is an evolutionary characteristic like having two ears, one nose, or having three cone cells and hence trichromatic vision. The brain seems to be formatted with a certain view of reality, that could be more or less inevitable due to our physical condition.

But it could be different. I always wondered how different we would be if we could see more then three dimensions. Would the mathematics we have developed be the same? Probably, not completely. For example, if we were capable of seeing in four dimensions, the role occupied by real analysis would probably have been taken by complex analysis. It is in some sense much simpler apart from the handicap that the complex functions of complex variable have four dimensional graphs, and so at any rate not easily handled by someone that has a problem in visualizing beyond three dimensions.

References:

[1] Jameson, K. A., Highnote, S. M., & Wasserman, L. M. (2001) Richer color experience in observers with multiple photopigment opsin genes, Psychonomic Bulletin & Review 8 (2). 244–261. doi:10.3758/BF03196159

[2] Mancuso et all (2009) Gene therapy for red–green color blindness in adult primates, Nature 461 (7265):784-787.

[3] Abbott, Edwin A. (1884), Flatland: A Romance of many dimensions, Dover, New York.

[4] Abbott, Edwin A. (2010), Flatland, an edition with notes and commentary by W. Lindgren and T. Banchoff, Cambridge University Press.

[5] Groh, Jennifer M. (2014) Making space - How the brain knows where things are, Belknap Press.

[6] Damásio, António (2000) The Feeling of What Happens: Body and Emotion in the Making of Consciousness, Mariner Books.

[7] L. W. Barsalou, Perceptual Symbol Systems, Behavioral and Brain Sciences 22 (04), 577-660.

[8] Alexander Schlegel, Peter J. Kohler, Sergey V. Fogelson, Prescott Alexander, Dedeepya Konuthula, and Peter Ulric Tse, Network structure and dynamics of the mental workspace, PNAS, 110 (40), 16277-16282.